\documentclass[secnumarabic,amssymb, nobibnotes, aps, prd, 11pt, abstract]{revtex4-2}
\usepackage{tocloft}
\usepackage[dvips]{graphicx}
\usepackage{amsmath}
\usepackage{subcaption}
\usepackage{hyperref}
\usepackage{filecontents}
\hypersetup{colorlinks,linkcolor={red},citecolor={blue},urlcolor={red}}
\usepackage{setspace}
\usepackage{xcolor}
\pagecolor{white}

\usepackage{multirow}
\color{black}

\begin{document}

\singlespacing

	\title{Complex Phase Structure and Widom line for Euler Heisenberg black holes}%

	\author{Mozib Bin Awal$^1$}
	
	\email{$rs_mozibbinawal@dibru.ac.in$}
	
	\author{Prabwal Phukon$^{1,2}$}
	\email{$prabwal@dibru.ac.in$}
	
	\affiliation{$^1$Department of Physics, Dibrugarh University, Dibrugarh, Assam,786004.\\$^2$Theoretical Physics Division, Centre for Atmospheric Studies, Dibrugarh University, Dibrugarh,Assam,786004.\\}

	\begin{abstract}
	
	We investigate the supercritical thermodynamics of Euler-Heisenberg AdS black holes within the framework of Lee-Yang phase transition theory. We show that the system admits two distinct critical points associated with a four-phase thermodynamic structure and identify a degenerate higher-order critical point where the two criticalities merge. Extending the thermodynamic description into the complex domain, we determine the distribution of Lee-Yang singularities and construct the corresponding complex phase diagrams. At the degenerate critical point, we find that a well-defined Widom line emerges despite the absence of a conventional coexistence curve, acting as an effective stability boundary in the supercritical regime. In the two-critical-point regime, the complex phase diagram exhibits two distinct Widom lines, one associated with a coexistence curve and the other arising solely from the complex singularity structure. We further show that the Lee-Yang formalism consistently reproduces the expected phase structure for systems with a single critical point and in the absence of criticality. Our results reveal a rich supercritical phase structure and provide new insights into the origin and physical interpretation of Widom lines.

\end{abstract}
	
	\maketitle
	
\section{Introduction}\label{sec1}
Ever since the pioneering developments of black hole thermodynamics by Bekenstein, Hawking, and others in the 1970s \cite{Phys,bekens,Hawking,Hawking2,Bardeen}, the subject has remained one of the most active and fascinating areas of research in theoretical physics. The realization that black holes possess thermodynamic properties such as temperature and entropy established a profound connection between gravitation, quantum theory, and statistical mechanics. Following this development, several fundamental questions naturally emerged regarding the thermodynamic behaviour of black holes and their correspondence with conventional thermodynamic systems. Among these questions, a particularly intriguing one concerned the possibility of phase transitions in black holes, analogous to those observed in ordinary thermodynamic substances. Early investigations in this direction were carried out by Davies \cite{Davies} and Hut \cite{Hut}, who studied thermodynamic instabilities and phase-transition-like phenomena in black hole spacetimes. Subsequently, the advent of the AdS/CFT correspondence, proposed by Maldacena, provided a powerful framework linking gravitational theories in asymptotically anti-de Sitter (AdS) spacetimes to conformal field theories on their boundaries. This correspondence significantly enhanced interest in the thermodynamics of AdS black holes, as their thermal behavior could be interpreted in terms of phase structures of the dual field theories.

Around the same period, a novel perspective on black hole thermodynamics began to emerge in which the cosmological constant, $\Lambda$, was promoted to a thermodynamic variable and identified with the pressure of the system. Incorporating this pressure term into the first law of black hole thermodynamics led to the formulation of extended phase space thermodynamics, often referred to as black hole chemistry. This framework not only enriched the thermodynamic description of black holes but also revealed striking analogies with classical thermodynamic systems, including the celebrated Van der Waals liquid-gas phase transition \cite{Kubiz,Hawkpage,Cai,Kastor,Dolan,Dolan2,Dolan3,Kubizna,Xu,Xu2,Zhang}. 

For many years, investigations of black hole thermodynamics have primarily focused on phenomena occurring below the critical point, where universal critical behaviour and various thermodynamic phase transitions can be observed. Most known phase transitions terminate within this region. Nevertheless, a comprehensive understanding of black hole thermodynamics requires extending the analysis into the supercritical regime. In ordinary thermodynamic systems, extensive studies of substances such as water and carbon dioxide have revealed a variety of intriguing properties beyond the critical point. For systems possessing a single critical point, such as the van der Waals fluid, the supercritical region is commonly divided into liquid-like and gas-like sectors by the Widom line. Emerging from the critical point, this line is usually identified through extrema of thermodynamic response functions, such as the specific heat, and is regarded as a smooth continuation of critical behaviour into the supercritical domain \cite{simeoni2010widom,10.1063/1.4930542,NISHIKAWA1995149,doi:10.1021/acs.jpcb.9b04058,li2024thermodynamic,G.O.Jones_1956,xu2005relation,PhysRevE.86.052103,PhysRevLett.112.135701,PhysRevE.95.052120}. On the other hand, systems characterized by multiple critical points or triple points may exhibit significantly richer supercritical structures, including the presence of multiple Widom lines and several distinct crossover regions \cite{li2024thermodynamic}.

Recently, increasing attention has been directed toward understanding the supercritical properties of black hole systems. In this context, the Lee-Yang theory of phase transitions provides a powerful framework for investigating critical phenomena from a statistical-mechanical perspective \cite{Yang:1952be,Lee:1952ig}. The central idea of this theory is that phase transitions are encoded in the distribution of partition-function zeros in the complex plane, which correspond to singularities of the Gibbs free energy. Early studies explored thermodynamic geometric aspects of AdS black holes above the critical point \cite{Sahay:2017hlq}, while subsequent works examined supercritical behaviour and related dynamical characteristics in a variety of black hole backgrounds \cite{Zhao:2025ecg,Xu:2025jrk,Wang:2025ctk}. More recently, the Lee-Yang formalism has been employed to construct complex phase diagrams for charged AdS black holes, leading to the identification of a Widom line that separates the supercritical region into small-black-hole-like and large-black-hole-like phases \cite{Xu:2025jrk}. Following these developments, several studies extended the Lee-Yang framework to black hole systems exhibiting more intricate thermodynamic phase structures. In particular, the supercritical behaviour and complex phase diagrams of five- and six-dimensional charged Gauss-Bonnet AdS black holes were investigated in \cite{gb}, where the emergence of multiple Widom lines was shown to be closely related to the existence of triple-point phase structures. Subsequently, the complex phase diagram and supercritical crossover phenomena of Born-Infeld AdS black holes were explored in \cite{bi}, demonstrating that the Lee-Yang formalism remains applicable even in the presence of nonlinear electrodynamic effects and reentrant phase transitions. More recently, the authors of \cite{Anand} analyzed the Widom line and supercritical crossover behaviour in noncommutative-corrected black hole spacetimes. Interestingly, they reported the emergence of a scaling behaviour consistent with the mean-field universality class, providing further evidence for the universal nature of supercritical phenomena in gravitational thermodynamic systems.

In the present work, we investigate the complex phase structure and supercritical behaviour of Euler-Heisenberg AdS black holes, a class of black hole solutions arising from the coupling of gravity to nonlinear electrodynamics. The thermodynamics of Euler-Heisenberg AdS black holes has attracted considerable attention in recent years \cite{Magos}. A key feature of these systems is the dependence of their phase structure on the Euler-Heisenberg parameter $a$. Different ranges of $a$ give rise to distinct thermodynamic behaviors. When $a<0$, the black hole exhibits a standard first-order transition between small and large black hole phases, closely resembling the van der Waals liquid--gas transition. In the interval $0\leq a\leq \frac{32}{7}Q^2$, the system develops a reentrant phase transition. However, for sufficiently large values of the parameter, namely $a>\frac{32}{7}Q^2$, the phase transition disappears and only a single thermodynamic phase remains \cite{Ye}.

The remainder of this paper is organized as follows. In section \ref{sec2}, we review the thermodynamic properties and critical behaviour of the Euler--Heisenberg AdS black hole. In section \ref{sec3}, we employ the Lee-Yang formalism to construct the complex phase diagrams and investigate the associated Widom-line structures for the degenerate-critical-point, two-critical-point, single-critical-point, and no-criticality regimes. Finally, in section \ref{sec4}, we summarize our main findings and discuss their physical implications.

\section{Euler-Heisenberg AdS Black Holes}\label{sec2}
We start from the action of the Euler-Heisenberg black hole, given by \cite{Ple,Gogoi,Sal}
\begin{equation}\label{eq1}
S=\frac{1}{4\pi}\int_{M^4} d^4x \sqrt{-g}\left[\frac{1}{4}(R-2\Lambda)-
\mathcal{L}(F,G) \right]
\end{equation}
In the above action, $g$ denotes the determinant of the metric tensor and $R$ is the Ricci scalar. The quantity $\mathcal{L}(F,G)$ represents the nonlinear electrodynamics Lagrangian, formulated in terms of the electromagnetic invariants $F=\frac{1}{4}F_{\mu\nu}F^{\mu\nu}$, and $G=\frac{1}{4}F_{\mu\nu}\,{}^{*}F^{\mu\nu}$,
with $F_{\mu\nu}$ being the electromagnetic field-strength. In the case of Euler--Heisenberg electrodynamics, the Lagrangian density takes the form
\begin{equation}\label{eq2}
\mathcal{L}(F,G)=-F+\frac{a}{2}F^2+ \frac{7a}{8} G^2
\end{equation}
where $a$ determines the strength nonlinear contributions and it is known as the Euler-Heisenberg parameter. The line element for this particular black hole system is given as follows
\begin{equation}\label{eq3}
ds^2=f(r)dt^2+\frac{dr^2}{f(r)}+r^2(d\theta^2+\sin^2\theta d\phi^2)
\end{equation}
with the following lapse function
\begin{equation}\label{eq4}
f(r)= 1-\frac{2M}{r}+\frac{Q^2}{r^2}+\frac{ r^2}{l^2}-\frac{a Q^4}{20 r^6},
\end{equation}
With the above information, we can then easily determine the necessary thermodynamic quantities. The mass and temperature in terms of the horizon radius $r_+$ are found to be
\begin{equation}\label{eq5}
M=\frac{r_+}{2}+\frac{Q^2}{2 r_+}+\frac{4}{3} \pi  P r_+^3-\frac{a Q^4}{40 r_+^5}
\end{equation}
\begin{equation}\label{eq6}
T=\frac{a Q^4+32 \pi  P r_+^8-4 Q^2 r_+^4+4 r_+^6}{16 \pi  r_+^7}
\end{equation}
Using the relation $F=M-TS$, we can then get the expression of the Gibbs free energy as follows,
\begin{equation}\label{eq7}
F=\frac{3 Q^2}{4 r_+}+\frac{r_+}{4}-\frac{7 a Q^4}{80 r_+^5}-\frac{2}{3} \pi  P r_+^3
\end{equation}
where the entropy $S=\pi r_+^2$ is used. Lastly, we may also calculate the heat capacity $C_P$ using $C_P=T\left(\frac{\partial S}{\partial T}\right)_P $, which comes out to be
\begin{equation}\label{eq8}
C_P=\frac{2 \pi  r_+^2 \left(a Q^4+4 \left(8 \pi  P r_+^8-Q^2 r_+^4+r_+^6\right)\right)}{-7 a Q^4+32 \pi  P r_+^8+12 Q^2 r_+^4-4 r_+^6}
\end{equation}
\begin{figure}[htbp]
    \centering
    \begin{subfigure}[b]{0.45\textwidth}
        \includegraphics[width=\textwidth]{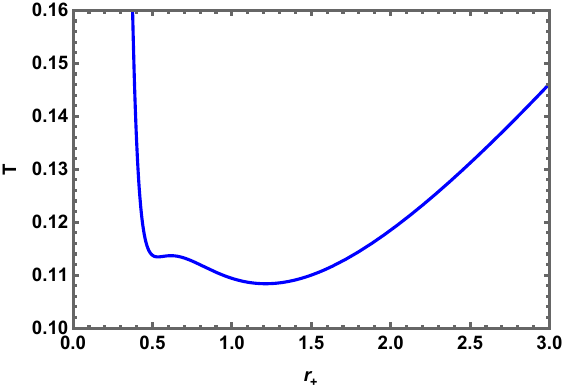}
        \caption{}
        \label{f1a}
    \end{subfigure}
    \hspace{0.05\textwidth}
    \begin{subfigure}[b]{0.45\textwidth}
        \includegraphics[width=\textwidth]{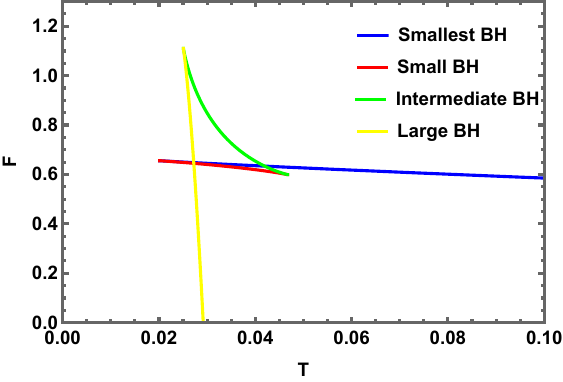}
        \caption{}
        \label{f1b}
    \end{subfigure}
    
    \vspace{0.5cm} 

    \begin{subfigure}[b]{0.45\textwidth}
        \includegraphics[width=\textwidth]{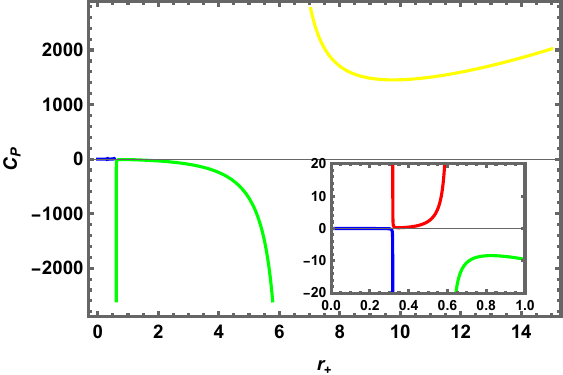}
        \caption{}
        \label{f1c}
    \end{subfigure}
  
    \caption{Thermodynamic behaviour of the Euler-Heisenberg AdS black hole in the four-phase regime. (a) Hawking temperature as a function of the horizon radius. (b) Gibbs free energy as a function of temperature. (c) Specific heat as a function of the horizon radius. Here the same colour coding is used as in Fig. 1b for different BH branch.}
    \label{f1}
\end{figure}

Interestingly, for the Euler--Heisenberg AdS black hole, solving the criticality conditions yields two distinct sets of critical parameters. The corresponding critical values are given by
\begin{equation}\label{eq9}
\frac{\partial T}{\partial r_+}=0,\quad \frac{\partial^2 T }{\partial r^2_+}=0
\end{equation}
Interestingly, for the Euler-Heisenberg black holes, we find two set of critical values after the solving the above equations. The obtained critical values are
\begin{equation}\label{eq10}
r_{+c1}=0.58989,\quad Q_{c1}=0.41467,\quad T_{c1}=0.07888
\end{equation}
and
\begin{equation}\label{eq11}
r_{+c2}=4.45590,\quad Q_{c2}=1.82270,\quad T_{c2}=0.02378
\end{equation}
where, throughout this analysis, the pressure is fixed at $P=0.001$. Below the critical value, the thermodynamic behaviour of the Euler-Heisenberg black hole becomes remarkably rich, exhibiting four distinct black hole phases. This feature is evident from the behavior of the Hawking temperature as a function of the horizon radius, shown in Fig.~\ref{f1a}. The temperature profile possesses three extrema, which partition the parameter space into four separate branches corresponding to different black hole phases. The existence of these four phases is further illustrated in the Gibbs free energy-temperature diagram presented in Fig.~\ref{f1b}. Following the terminology adopted in this work, these branches are referred to as the Smallest Black Hole (SmBH), Small Black Hole (SBH), Intermediate Black Hole (IBH), and Large Black Hole (LBH) phases, respectively, and are distinguished by different color schemes in the figure. Additional insight into the local thermodynamic stability of these phases is provided by the behaviour of the specific heat, plotted as a function of the horizon radius in Fig.~\ref{f1c}. The specific heat exhibits three divergences, signalling the boundaries between adjacent branches. As usual, phases with positive specific heat are locally thermodynamically stable, whereas those characterized by negative specific heat are unstable.

\begin{figure}[h!]
    \centering
    \begin{subfigure}[b]{0.45\textwidth}
        \includegraphics[width=\textwidth]{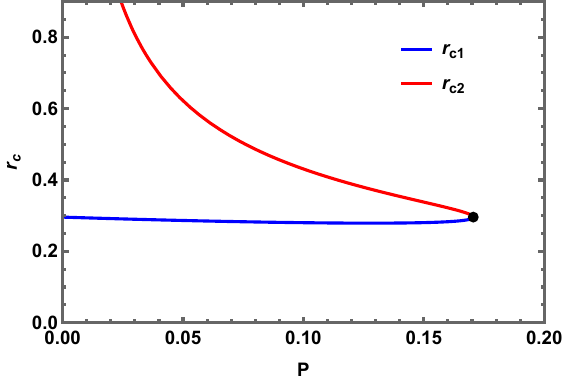}
        \caption{}
        \label{f2a}
    \end{subfigure}
    \hspace{0.05\textwidth}
    \begin{subfigure}[b]{0.45\textwidth}
        \includegraphics[width=\textwidth]{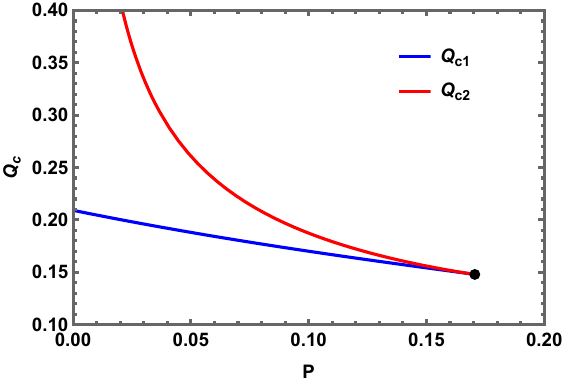}
        \caption{}
        \label{f2b}
    \end{subfigure}
    \caption{Merger of the two distinct critical points at the degenerate critical point}
    \label{f2}
\end{figure}

The existence of two distinct critical points naturally raises the question of whether these criticalities can merge for a particular choice of the thermodynamic parameters. To investigate this possibility, we impose the stronger conditions

\begin{equation}
\left(\frac{\partial T}{\partial r_+}\right)=
\left(\frac{\partial^2 T}{\partial r_+^2}\right)=
\left(\frac{\partial^3 T}{\partial r_+^3}\right)=0,
\end{equation}
which correspond to the simultaneous vanishing of the first three derivatives of the Hawking temperature with respect to the horizon radius. Solving these equations yields a unique set of values $(r_{+d}=0.29580, Q_d=0.14790,P_d=0.17052)$, indicating the coalescence of the two ordinary critical points into a single degenerate critical point.The merger of the two critical points can be visualized by tracking the solutions of Eq.~\ref{eq9} as a function of the pressure, as illustrated in Fig.~\ref{f2}. With increasing pressure, the two initially distinct critical points continuously move towards each other and ultimately coalesce into a single point, marked by the black dot in Figs.~\ref{f2a} and \ref{f2b}. This point represents the degenerate critical point $(r_{+d},,Q_d,,P_d)$, at which the two ordinary critical points cease to exist independently and merge into a single higher-order critical point.

\begin{figure}[t!]
    \centering
    \begin{subfigure}[b]{0.45\textwidth}
        \includegraphics[width=\textwidth]{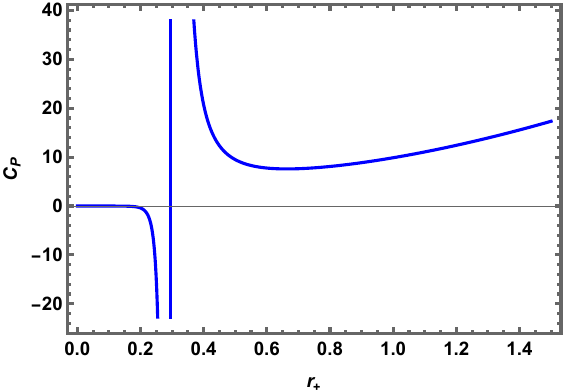}
        \caption{Specific heat}
        \label{f3a}
    \end{subfigure}
    \hspace{0.05\textwidth}
    \begin{subfigure}[b]{0.45\textwidth}
        \includegraphics[width=\textwidth]{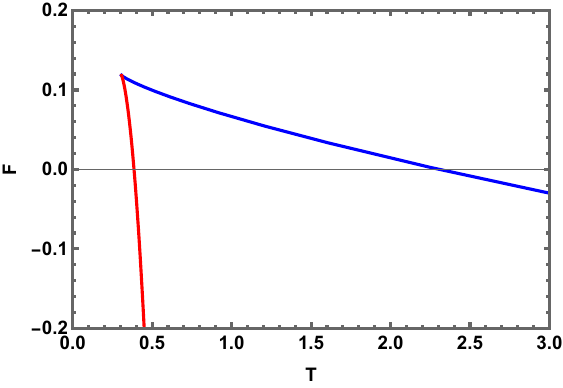}
        \caption{Free energy as a function of temperature}
        \label{f3b}
    \end{subfigure}
    \caption{The two phases of the Euler-Heisenberg AdS black hole at the degenerate critical point.}
    \label{f3}
\end{figure}

At this special point, the temperature profile develops a higher-order stationary inflection point, signalling a qualitative change in the phase structure of the system. In contrast to the generic situation where two distinct critical points bound the four-phase region, the degenerate point marks the threshold at which the corresponding criticalities become indistinguishable. The Gibbs free energy at the degenerate critical point is displayed in Fig.~\ref{f3b}. One observes that the separate free-energy branches associated with the multiple black-hole phases merge into a simplified structure, reflecting the disappearance of the distinction between the two critical behaviours. This behaviour provides strong evidence that the degenerate point acts as the merger point of the two criticalities and represents a higher-order critical phenomenon in the thermodynamics of the Euler-Heisenberg black hole. 
The heat capacity profile with respect to horizon radius is shown in Fig.~\ref{f3a} which shows that among the two branches one has negative specific heat signifying that it is the unstable branch and the one with the positive specific heat is the stable branch.

\section{Complex Phase Diagram and Widom Line}\label{sec3}
\subsection{Two-Critical-Point and Degenerate-Critical-Point Regimes}
We now extend our analysis of the Euler-Heisenberg AdS black hole into the supercritical regime by investigating its thermodynamic structure in the complex domain. To achieve this, we employ the Lee-Yang theory of phase transitions, which provides a powerful framework for understanding critical phenomena through the distribution of singularities in the complex plane.

The central idea of Lee-Yang theory is that the thermodynamic properties of a system are encoded in the zeros of its partition function. In the thermodynamic limit, a phase transition occurs when these complex zeros approach and eventually accumulate on the real axis, giving rise to non-analytic behaviour in the thermodynamic potentials. Within the Euclidean formulation of black hole thermodynamics, the partition function is related to the Euclidean action through
\begin{equation}
Z \sim e^{-I_E}
\end{equation}
while the Gibbs free energy is given by
\begin{equation}
G = T I_E
\end{equation}
Combining these relations yields
\begin{equation}
G=-T\ln Z
\end{equation}
indicating that the zeros of the partition function correspond to singularities of the Gibbs free energy.

For the Euler-Heisenberg black hole, such singularities are manifested through divergences of the specific heat. Since the specific heat is proportional to the second derivative of the Gibbs free energy, the locations of its divergences identify the Lee-Yang singularities of the system. Consequently, determining the Lee-Yang zeros reduces to solving the condition that the denominator of the specific heat vanishes. While these singularities lie on the real axis in the critical region, they generally move into the complex plane once the system enters the supercritical regime.

It is important to emphasize that the horizon radius is not assumed to acquire a physical complex value. Rather, the extension to the complex plane should be viewed as an analytic continuation of the thermodynamic functions, employed as a mathematical tool to probe the underlying singularity structure of the system. In this framework, the disappearance of a first-order phase transition beyond the critical point is associated with the migration of the corresponding singularities away from the real axis. The trajectories of these complex singularities can then be projected onto the physical thermodynamic plane, providing a natural definition of the Widom line and the associated supercritical crossover. Unlike the Reissner-Nordström AdS black hole, the Euler-Heisenberg system possesses two distinct critical points which then merge at the double critical point. It is therefore natural to investigate how the associated Lee-Yang singularities evolve in the supercritical regime and whether the resulting Widom-line structure reflects the richer thermodynamic landscape of the system.
\begin{figure}[h!]
    \centering
    \begin{subfigure}[b]{0.45\textwidth}
        \includegraphics[width=\textwidth]{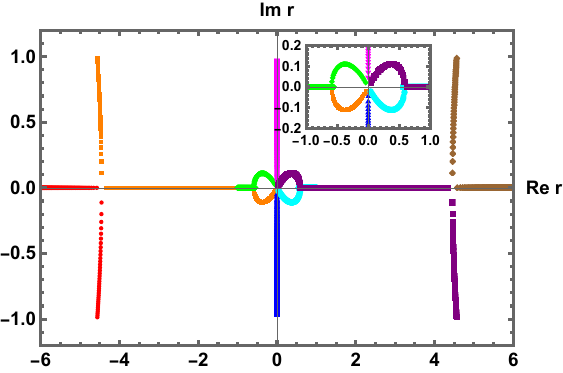}
        \caption{}
        \label{f4a}
    \end{subfigure}
    \hspace{0.05\textwidth}
    \begin{subfigure}[b]{0.45\textwidth}
        \includegraphics[width=\textwidth]{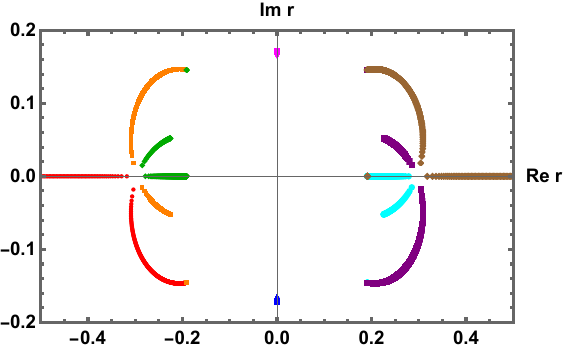}
        \caption{}
        \label{f4b}
    \end{subfigure}
    \caption{Distribution of the Lee-Yang singularities in the complex horizon-radius plane. Panel (a) corresponds to the regime with two distinct critical points, while panel (b) shows the singularity distribution at the degenerate point where the two criticalities merge.}
    \label{f4}
\end{figure}
Figure~\ref{f4} illustrates the evolution of the Lee-Yang singularities in the complex horizon radius plane. The left panel corresponds to a parameter regime in which two distinct critical points exist. In this case, the singularities organize themselves into two separate families of trajectories, each associated with one of the critical points. The points lying on the real axis correspond to the spinodal boundaries, while the remaining singularities appear as complex-conjugate pairs that move away from the real axis upon entering the supercritical regime. The trajectories of the Lee-Yang singularities intersect at the real axis at exactly the critical points (at $0.58989$ and $4.45590$). The inset highlights the intricate structure of the singularities in the vicinity of the origin, where the two branches remain clearly distinguishable.

As the system approaches the critical-point merger, the topology of the singularity structure changes significantly, as shown in the right panel. The two previously independent families of Lee-Yang singularities move towards each other and eventually coalesce into a single configuration. A particularly noteworthy feature is that the branches emerging from the positive and negative imaginary axes intersect the real axis at the same point, signalling the merger of the two critical structures. The point of intersection is again exactly the degenerate critical point ($0.29580$). This behaviour suggests the formation of a higher-order critical configuration and marks the transition from a two-critical-point phase structure to a single degenerate critical point.
\begin{figure}[h!]
\centering
\includegraphics[width=0.97\textwidth]{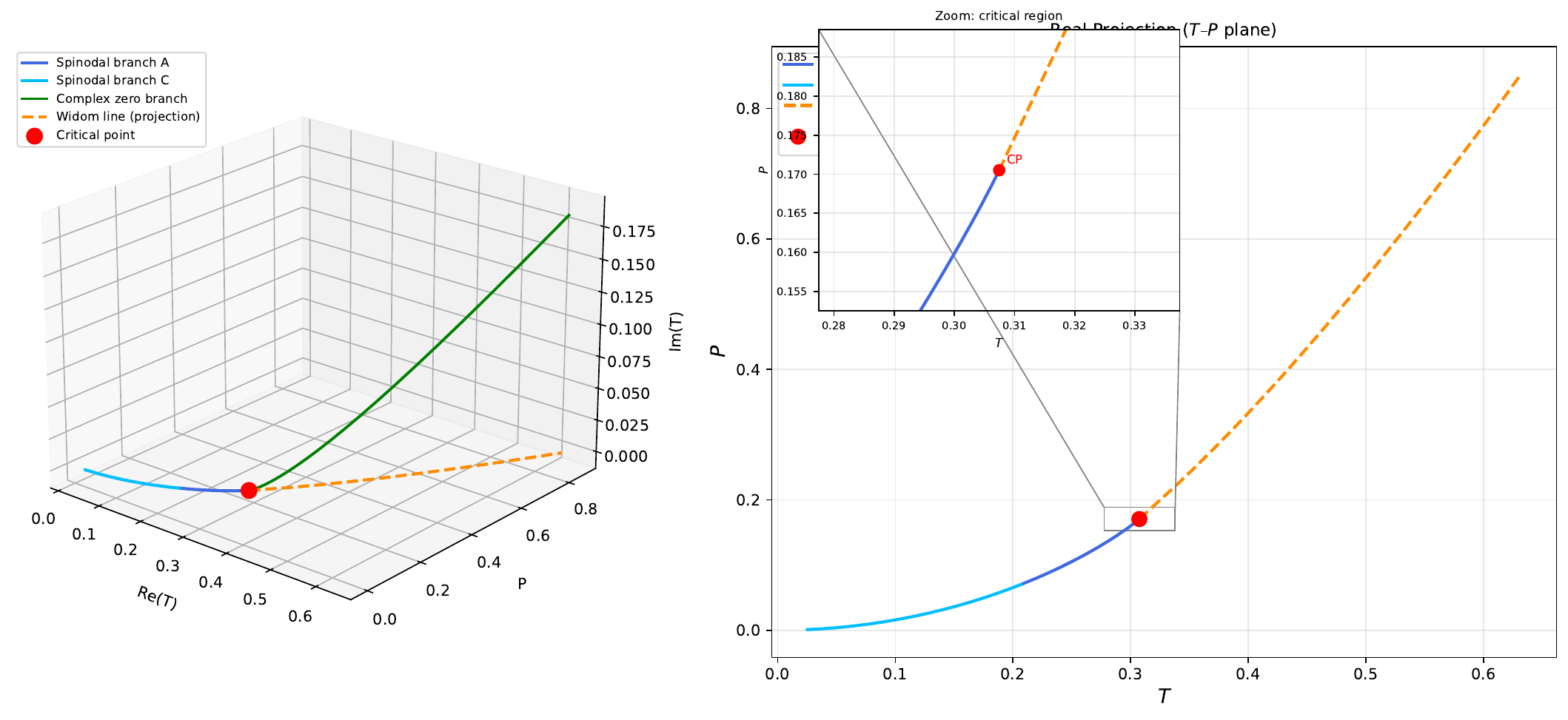}
\caption{Complex phase diagram of the Euler-Heisenberg AdS black hole at the degenerate critical point. The left panel shows the complete three-dimensional phase structure, while the right panel shows its projection onto the $P$--$T$ plane.}
\label{f5}
\end{figure}
The complete complex phase diagram of the Euler-Heisenberg black hole is presented in Fig.~\ref{f5}, where all parameters are fixed at their corresponding values at the degenerate critical point. Extending the thermodynamic phase structure into the complex domain reveals several features that are absent in the conventional real-valued phase diagram. In particular, two additional crossover curves emerge, namely the complex line and the Widom line, represented by the green and orange curves, respectively, in Fig.~\ref{f5}. These lines originate from the complex singularity structure of the Gibbs free energy and characterize the evolution of the thermodynamic system beyond the critical region. Consequently, the complex phase diagram provides a more complete description of the phase structure, incorporating both the conventional critical behaviour on the real plane and the supercritical crossover phenomena encoded in the complex domain.

The complex phase line is constructed from the complex singularities of the Gibbs free energy obtained from the divergence condition of the specific heat. Following the Lee-Yang prescription, we focus on the singularities located in the first quadrant of the complex horizon-radius plane and determine their corresponding thermodynamic coordinates through the equation of state. The resulting trajectory defines the complex phase line shown in Fig.~\ref{f5}. Unlike conventional phase boundaries, this line resides in the complexified thermodynamic space and encodes information about the analytic structure of the system beyond the critical region. The physically relevant information is obtained by projecting the complex phase line onto the real thermodynamic plane. This projection defines the Widom line, which extends into the supercritical region and characterizes the thermodynamic crossover beyond the critical point. In this sense, the Widom line may be regarded as the real-plane manifestation of the underlying complex singularity structure, providing a natural boundary between different supercritical regimes without involving any thermodynamic singularity.

Let us now examine the phase structure of the Euler-Heisenberg AdS black hole in greater detail. The light-blue and dark-blue curves in Fig.~\ref{f5} represent the two spinodal lines, obtained from the loci of the real positive singularities of the Gibbs free energy. These singularities correspond to the divergence points of the specific heat and mark the boundaries between different thermodynamic branches.

A notable feature of the phase diagram is the absence of a coexistence line terminating at the degenerate critical point. This behaviour is consistent with the thermodynamic structure discussed in the previous section. As shown in Fig.~\ref{f3}, at the degenerate critical point the system consists of only two branches, one locally stable with positive specific heat and the other unstable with negative specific heat. Since a first-order phase transition requires the coexistence of two stable phases with equal Gibbs free energy, no conventional coexistence curve emerges in this case.

As a consequence, the physical interpretation of the Widom line differs from that encountered in standard van der Waals-like black hole systems. There, the Widom line is generally viewed as the continuation of the coexistence curve into the supercritical region and therefore separates small-black-hole-like and large-black-hole-like phases. In contrast, for the Euler-Heisenberg black hole the Widom line cannot be associated with the continuation of a first-order coexistence boundary. Instead, it should be regarded as a crossover line originating from the underlying complex singularity structure of the Gibbs free energy, separating distinct supercritical regimes without being directly connected to a conventional SBH/LBH phase transition.

\begin{figure}[h!]
\centering
\includegraphics[width=0.5\textwidth]{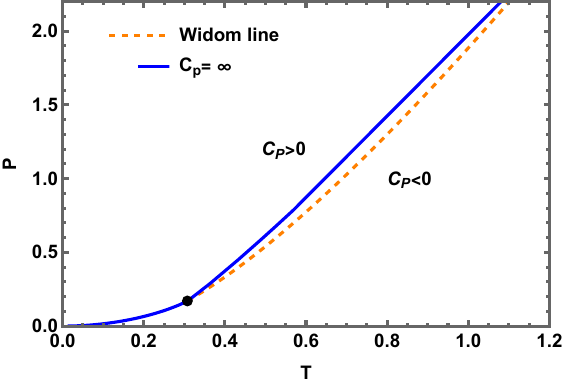}
\caption{FIG. 6: Comparison between the Widom line and the spinodal line ($C_P=\infty$) for the Euler-Heisenberg AdS black hole. The two curves nearly coincide, indicating that the Widom line effectively tracks the spinodal boundary in the supercritical regime.}
\label{f6}
\end{figure}

Remarkably, although the Euler--Heisenberg AdS black hole does not exhibit a conventional coexistence curve in the real thermodynamic parameter space, the complexified analysis reveals a well-defined branch of Lee-Yang zeros emanating from the critical point. We find that the projection of this complex branch onto the real $(T, P)$ plane nearly coincides with the locus determined by the divergence condition of the specific heat as observed in Fig.~\ref{f6}. This striking agreement indicates that the Lee-Yang branch represents the complex continuation of the spinodal structure beyond the critical point. Furthermore, an explicit evaluation of the specific heat on either side of the projected curve reveals a change in sign, with $C_P>0$ on one side and $C_P<0$ on the other (Fig.~\ref{f6}). Consequently, the projected Lee-Yang branch acts as an effective stability boundary separating thermodynamically stable and unstable sectors of the parameter space, despite the absence of a conventional first-order coexistence line. These results suggest that the underlying critical structure of the Euler-Heisenberg AdS black hole persists in the complex thermodynamic manifold and manifests itself through a complex spinodal boundary whose real projection retains direct thermodynamic significance.

\begin{figure}[h!]
\centering
\includegraphics[width=0.97\textwidth]{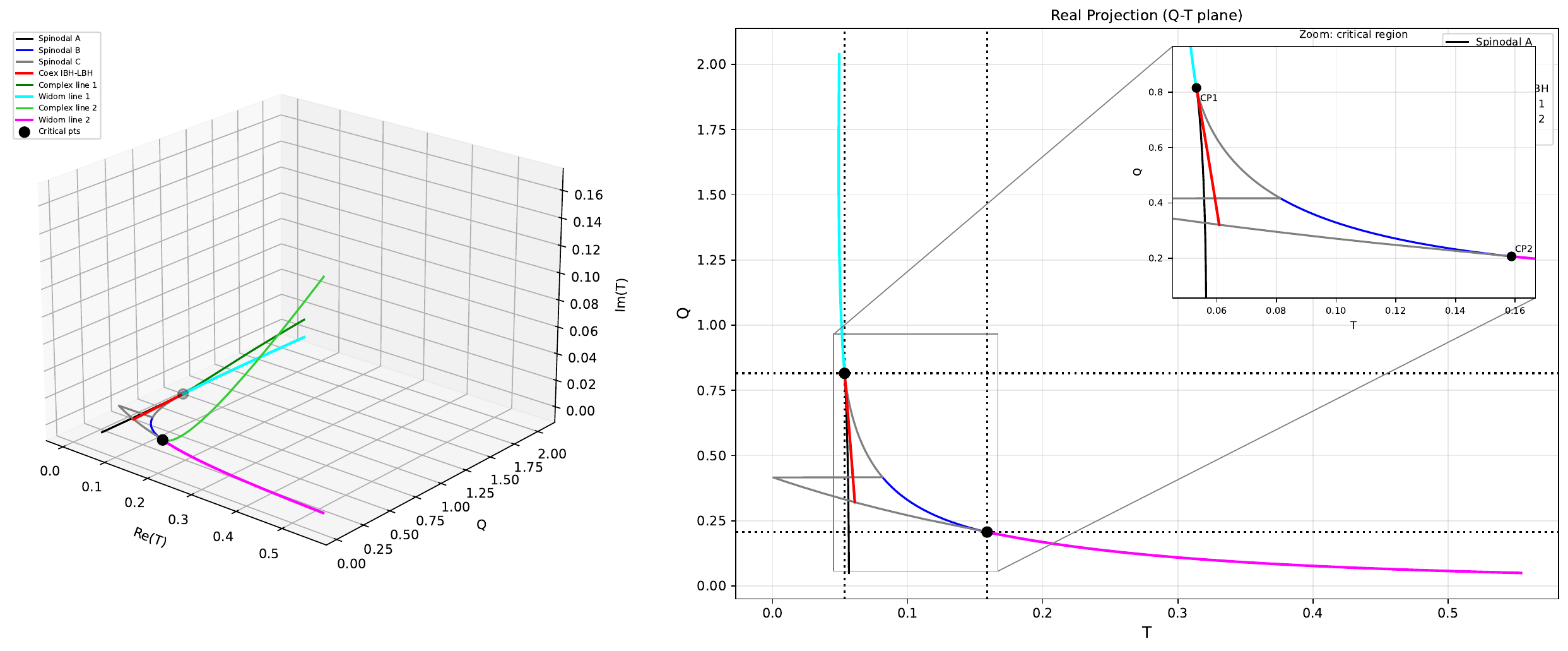}
\caption{Complex phase diagram of the Euler-Heisenberg AdS black hole in the two-critical-point regime. The left panel shows the complete three-dimensional phase structure, while the right panel shows its projection onto the $P$--$T$ plane.}
\label{f7}
\end{figure}

Next we study the complex phase diagram for the case when we have two critical points, for which the distribution of the Lee-Yang singularities in the complex plane is shown in Fig.~\ref{f4a}, while the associated Gibbs free energy and specific heat profiles are shown in Fig.~\ref{f1}. As discussed previously, the specific heat exhibits four distinct thermodynamic branches, two of which are locally stable with positive heat capacity and two of which are unstable with negative heat capacity. The resulting complex phase diagram is presented in Fig.~\ref{f7}. A conventional coexistence line, shown in red, is observed to terminate at the first critical point (CP1). This coexistence curve separates the small and large black hole phases and is accompanied by a Widom line (cyan) that extends into the supercritical region. The Widom line is obtained as the projection of the corresponding complex phase line (dark green curve in the three-dimensional representation) onto the real thermodynamic plane. In analogy with other AdS black hole systems, this crossover line divides the supercritical region into small-black-hole-like and large-black-hole-like sectors.

In addition, the phase diagram contains three spinodal curves, shown in black, blue, and gray. These curves are determined from the loci of the real positive singularities of the Gibbs free energy and represent the boundaries at which the specific heat diverges. Together they delineate the various stable and unstable thermodynamic branches of the system.

A particularly interesting feature emerges at the second critical point (CP2). Unlike CP1, no coexistence line terminates at this critical point, indicating the absence of a conventional first-order phase transition. Nevertheless, the complex singularity structure still gives rise to a well-defined complex phase line (light green in the three-dimensional plot), whose projection onto the real thermodynamic plane generates a second Widom line, shown in magenta. This observation demonstrates that the existence of a Widom line is not necessarily tied to the continuation of a coexistence curve. Instead, it originates from the underlying complex singularity structure of the Gibbs free energy. Consequently, the magenta Widom line should be interpreted as a crossover boundary separating distinct thermodynamic regimes. Similar to the degenerate-critical-point case discussed earlier, it acts as an effective stability boundary between sectors characterized by different thermodynamic behaviour, despite the absence of an associated coexistence line.

The emergence of two qualitatively different Widom-line structures within the same black hole system highlights the richness of the Euler-Heisenberg phase diagram. While the first Widom line retains the conventional interpretation as the supercritical continuation of a first-order phase transition, the second arises solely from the complex singularity structure and therefore represents a genuinely new type of thermodynamic crossover.

\subsection{Single-Critical-Point and No-Criticality Regimes}
\begin{figure}[h!]
\centering
\includegraphics[width=0.5\textwidth]{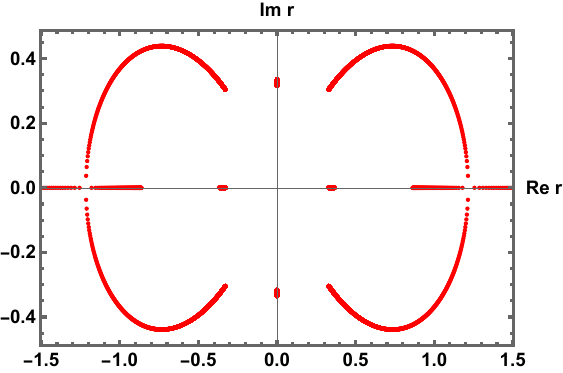}
\caption{Distribution of the Lee-Yang singularities in the complex horizon-radius plane corresponding to the regime with a single critical point.}
\label{f8}
\end{figure}

To further demonstrate the consistency of our analysis, we also examine the complex phase structure in two additional thermodynamic regimes: one in which the Euler-Heisenberg AdS black hole possesses a single critical point (distinct from the degenerate critical point) and another in which no criticality exists. We first consider the single-critical-point case, whose corresponding distribution of Lee-Yang singularities is shown in Fig.~\ref{f8}. The overall interpretation follows that discussed previously. The singularities lying on the real axis correspond to the spinodal boundaries, while the remaining singularities appear as complex-conjugate pairs that move away from the real axis in the supercritical regime. A characteristic feature of this configuration is that the trajectories of the Lee-Yang singularities intersect the real axis at a single location, corresponding to the unique critical point of the system. The resulting complex phase diagram therefore reproduces the conventional picture of a single critical point with one associated Widom line, demonstrating that the present formalism consistently captures both simple and more intricate thermodynamic phase structures.

\begin{figure}[h!]
\centering
\includegraphics[width=0.97\textwidth]{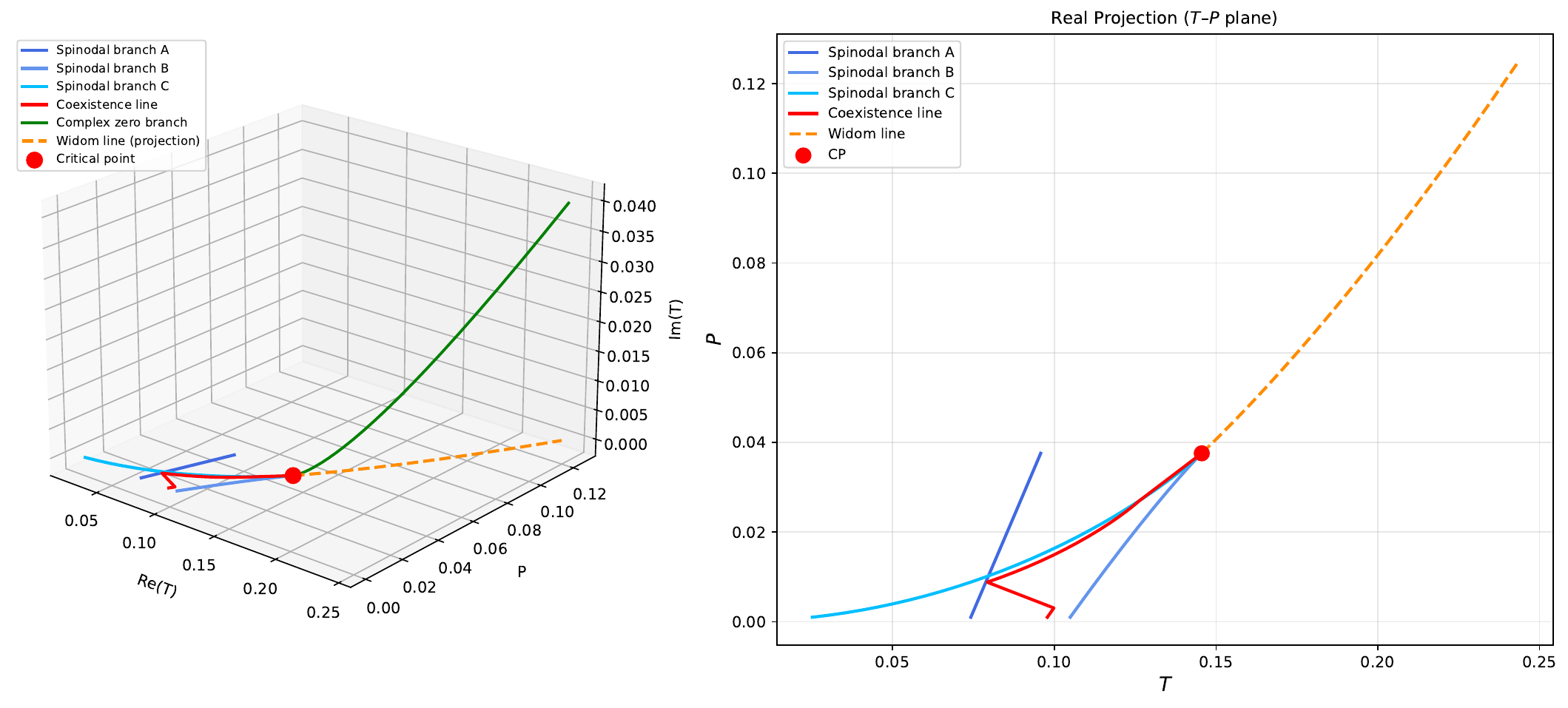}
\caption{Complex phase diagram of the Euler-Heisenberg AdS black hole at the single critical point. The left panel shows the complete three-dimensional phase structure, while the right panel shows its projection onto the $P$--$T$ plane.}
\label{f9}
\end{figure}

\begin{figure}[h!]
    \centering
    \begin{subfigure}[b]{0.45\textwidth}
        \includegraphics[width=\textwidth]{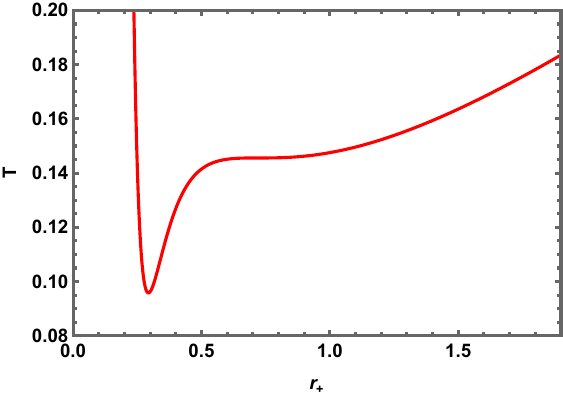}
        \caption{}
        \label{f10a}
    \end{subfigure}
    \hspace{0.05\textwidth}
    \begin{subfigure}[b]{0.45\textwidth}
        \includegraphics[width=\textwidth]{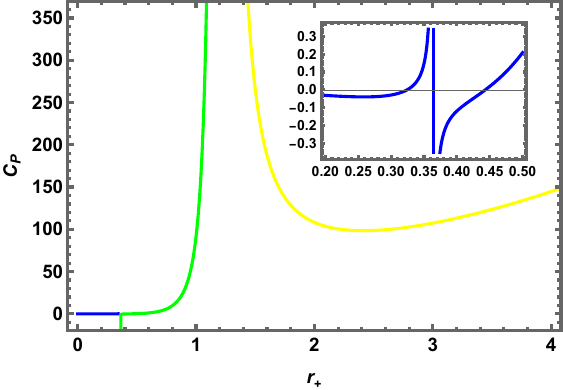}
        \caption{}
        \label{f10b}
    \end{subfigure}
    \caption{(a) Temperature and (b) specific heat profiles for Euler-Heisenberg black holes when only one critical point exists.}
    \label{f10}
\end{figure}
The corresponding complex phase diagram is presented in Fig.~\ref{f9}, where the critical point is indicated by the red dot. A single coexistence curve, shown in red, terminates at the critical point and separates the Intermediate Black Hole (IBH) and Large Black Hole (LBH) phases. This phase structure is consistent with the thermodynamic behaviour displayed in Fig.~\ref{f10}. In particular, the Hawking temperature profile shown in Fig.~\ref{f10a} possesses two extrema, while the specific heat plotted in Fig.~\ref{f10b} exhibits two locally stable branches with positive heat capacity, corresponding to the IBH and LBH phases, separated by an unstable branch with negative heat capacity.

Following the same procedure adopted in the previous cases, we construct the complex phase line from the trajectories of the Lee-Yang singularities in the complex thermodynamic space. The resulting complex phase line is shown by the green curve in the three-dimensional representation, whose projection onto the real thermodynamic plane gives rise to the Widom line (orange dashed curve). As expected, the Widom line originates at the critical point and extends into the supercritical region, separating the intermediate-black-hole-like and large-black-hole-like thermodynamic regimes. The spinodal boundaries, obtained from the loci of the real positive singularities of the Gibbs free energy, are represented by the light blue, sky blue, and royal blue curves.

\begin{figure}[h!]
\centering
\includegraphics[width=0.5\textwidth]{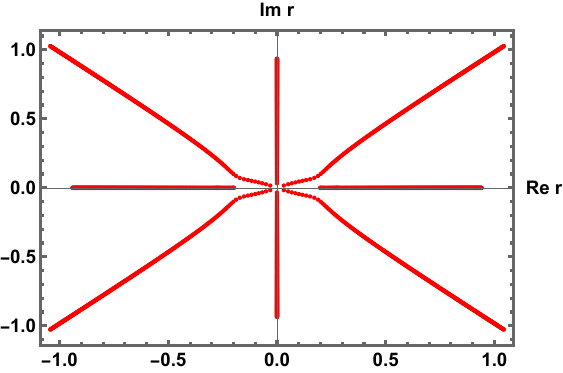}
\caption{Distribution of the Lee-Yang singularities in the complex horizon-radius plane corresponding to the regime with a no critical point.}
\label{f11}
\end{figure}

Finally, we consider the regime in which the Euler-Heisenberg AdS black hole exhibits no critical behaviour. For this purpose, the pressure and charge are chosen to be greater than their corresponding degenerate critical values. The resulting distribution of the Lee-Yang singularities in the complex horizon-radius plane is shown in Fig.~\ref{f11}. In contrast to the previous cases, the complex-conjugate trajectories remain detached from the real axis throughout their evolution and do not intersect it at any finite point. Consequently, no critical points are generated, in agreement with the absence of thermodynamic criticality in this parameter regime. This behaviour further illustrates that the approach of the Lee-Yang singularities towards the real axis is a characteristic signature of critical phenomena, whereas their failure to do so corresponds to a smooth thermodynamic landscape devoid of phase transitions.

\section{Conclusion and Discussion}\label{sec4}

In this work, we have investigated the supercritical thermodynamics of Euler-Heisenberg AdS black holes using the Lee-Yang theory of phase transitions. We first showed that the system possesses two distinct critical points associated with a four-phase thermodynamic structure consisting of the Smallest, Small, Intermediate, and Large black hole branches. By imposing the simultaneous vanishing of the first three derivatives of the Hawking temperature, we further identified a degenerate critical point at which the two ordinary criticalities merge into a higher-order critical configuration.

To explore the supercritical regime, we analytically continued the thermodynamic description into the complex domain and constructed the corresponding Lee-Yang singularity distributions. From these singularities, we obtained the complete complex phase diagrams and identified the associated complex phase lines and their real projections, namely the Widom lines. Our analysis shows that the complex singularity structure faithfully captures the evolution of the thermodynamic phase structure beyond the critical point.

For the degenerate critical point, we found that no conventional coexistence curve terminates at the critical point. Nevertheless, the complex singularity structure gives rise to a well-defined Widom line whose projection separates thermodynamically stable and unstable sectors of the supercritical region. This demonstrates that a physically meaningful Widom line can exist even in the absence of a first-order coexistence curve.

The situation becomes even richer in the two-critical-point regime. While one critical point exhibits the conventional scenario in which the Widom line is the supercritical continuation of a coexistence curve, the second critical point generates a distinct Widom line solely through the projection of the complex singularity structure, despite the absence of an associated coexistence boundary. The coexistence of these two qualitatively different Widom-line structures within the same thermodynamic system reveals that supercritical crossovers need not possess a unique geometric origin.

Finally, we examined the limiting cases in which the Euler-Heisenberg black hole possesses a single critical point and no criticality. In both situations, the Lee-Yang singularity distributions correctly reproduce the expected thermodynamic behaviour. The emergence of a single Widom line in the former case and the complete absence of critical signatures in the latter provide strong consistency checks of the present framework.

Overall, our results establish the complex phase diagram as a powerful tool for exploring thermodynamics of black hole in the supercritical regime . In particular, the emergence of Widom lines without accompanying coexistence curves suggests that complex singularities can encode physically relevant crossover phenomena even when conventional phase-transition signatures are absent. We expect that these findings may provide further insight into the role of Lee-Yang singularities in black hole thermodynamics and motivate similar investigations in other black hole systems exhibiting complicated, richer or more intricate phase structures.

\end{document}